\documentstyle[12pt]{article}
\def\rd{{\rm d}}
\begin{document}
\begin{titlepage}
\pagestyle{empty}
\baselineskip=21pt
\rightline{UMN-TH-1113-93}
\rightline{TPI-MINN-93/13-T}
\rightline{hep-ph/yymmddd}
\rightline{April 1993}
\vskip .2in
\begin{center}
{\large{\bf Constraints on Majoron Models, Neutrino Masses \\
 and Baryogenesis}}
\end{center}
\vskip .1in
\begin{center}
James M.~Cline

Kimmo Kainulainen

and

Keith A. Olive

{\it School of Physics and Astronomy, University of Minnesota}

{\it Minneapolis, MN 55455, USA}

\vskip .2in

\end{center}
\vskip 1in
\centerline{ {\bf Abstract} }
\baselineskip=18pt

\noindent
{\newcommand{\la}{\mbox{\raisebox{-.6ex}{~$\stackrel{<}{\sim}$~}}}
{\newcommand{\ga}{\mbox{\raisebox{-.6ex}{~$\stackrel{>}{\sim}$~}}}
We derive strong contraints on the Yukawa couplings and the vacuum expectation
value in the singlet majoron model, taking into account the possibility of a
small gravitationally induced mass for the majoron.  If the present baryon
asymmetry was created earlier than at the electroweak phase transition, then to
preserve it from the combined effects of lepton number violating interactions
and electroweak sphalerons, we find that stringent constraints on the Yukawa
couplings, $h < 10^{-7}$, and the majoron scale, $v_s < v_{EW}$, must be
satisfied.  We also carefully rederive baryogenesis bounds on neutrino masses,
finding that in general they apply not to the masses themselves, but only to
related parameters, and they are numerically somewhat less stringent than has
previously been claimed.
}}
\end{titlepage}
\baselineskip=18pt
{\newcommand{\la}{\mbox{\raisebox{-.6ex}{~$\stackrel{<}{\sim}$~}}}
{\newcommand{\ga}{\mbox{\raisebox{-.6ex}{~$\stackrel{>}{\sim}$~}}}
\def\lsh{\lambda_{s\scriptscriptstyle H}}
\def\lh{\lambda_{\scriptscriptstyle H}}
\def\Lt{\tilde L}
\section{Introduction}
There is an ever increasing interest in the possibility for neutrino
masses. Aside from the intrinsic interest in new and perhaps tangible physics,
neutrino masses have been shown to be quite useful. A neutrino with a small
mass, O(10 eV), could make a significant contribution to the overall energy
density of the Universe.   Other potential benefits from neutrino masses
include MSW neutrino mixing relating to the solar neutrino prolem \cite{msw},
and the possibility of exploiting lepton number violating interactions in
conjunction with baryon and lepton number violation in the electroweak sector
\cite{krs1}  to generate a baryon asymmetry of the Universe
\cite{fy1}.

The very small ratio of expected neutrino masses, relative to other particle
masses, made the idea of the seesaw mechanism quite attractive \cite{seesaw}.
Numerous models incorporate a neutrino mass seesaw, including
unification via SO(10) \cite{hrr}, or simply adding
a single complex singlet to the standard model which breaks a global
U(1)$_{B-L}$  symmetry, {\it ie.,} the singlet majoron model \cite{cmp}. In
both of these models lepton number is spontaneouly broken. In a GUT,
$(B-L)$-violating effects will generally be small because the breaking scale
is generically $M_{GUT}$. In the majoron model, there are {\it a priori}
few restrictions on the breaking scale, which might be close to the electroweak
scale.

However the possibility of lepton number violation at or above the
electroweak scale is potentially dangerous, as was first discussed by Fukugita
and Yanagida \cite{fy2} in the context of a simple model with explicit lepton
number violation. The danger occurs when $L$-violating interactions
are in equilibrium in the early Universe at the same time that
$(B+L)$-violating nonperturbative electroweak interactions, due to sphalerons,
are in equilibrium. In this case the baryon asymmetry of the Universe is
destroyed independently of the initial $B-L$ asymmetry. This realization has
led
to a series of contraints on models with additional sources of baryon or lepton
number violation
\cite{fy2}-\cite{cdeo3}.
These bounds are of
course evaded if the baryon asymmetry is ultimately produced during or after
the electroweak phase transition.

The baryon wash-out bounds also directly apply to the majoron model
if the lepton symmetry breaking occurs above the electroweak scale.
They take on an especially interesting twist however, when
combined with the ``expectation" of gravitational corrections to the
singlet sector. Recalling the well-known adage that Nature abhors
global symmetries, it is quite reasonable to expect nonrenormalizable
contributions to the low energy Lagrangian due to  gravitational corrections
that violate nongauged symmetries \cite{hpp}. Although such terms
explicitly break global symmetries, their effects are strongly felt
only above some cutoff, in this case the Planck scale.
Such effects were previously considered in the triplet
majoron model \cite{mas} and for axions \cite{bs}.
Nonrenormalizable gravitationally-induced
interactions which explicitly break the global
$B-L$ symmetry were recently studied in the context of the singlet majoron
model \cite{ak,sek}.
In \cite{ak} it was shown that these effects generate a mass O(1-2) keV or
larger for the majoron and that the cosmological consequences of such a massive
majoron imply a stringent upper bound on the vacuum expectation value of the
singlet field in the broken phase.

In this paper we derive some of the baryon wash-out bounds in detail. In
particular we extend the bound coming from decays and inverse decays
\cite{sonia} of heavy neutrinos to the case of spontaneously broken symmetry
and find an explicit expression for this bound in terms of the light neutrino
seesaw masses, the neutrino mixing matrix and the initial lepton asymmetries.
We will then show how the bound on the singlet vacuum expectation value may be
further strengthened when combined with the $(B+L)$-violating nonperturbative
electroweak effects.

\section{The Model}

The singlet majoron model extends the Standard Model by merely adding a
single complex scalar field, $\sigma$, and a two-component fermion field
$\nu_R$, both SU(2)$\times$U(1)  singlets. The new interaction
Lagrangian terms may be written as
\begin{equation}
h H {\bar L} \nu_R  +  h_2 \sigma {\bar \nu_R} {\nu_R}^c + h.c. +
V(\sigma,H);
\end{equation}
\begin{equation}
V(\sigma,H) = \lambda_s\left(|\sigma|^2-v_s^2/2\right)^2 +
\lh\left(|H|^2-v^2/2\right)^2 + \lsh(|\sigma|^2-v_s^2/2)(|H|^2-v^2/2),
\label{effpot}
\end{equation}
where we allow for self-interactions of the scalar singlet as well as
interactions with the Standard Model Higgs doublet, $H$. $L$ is the usual
left-handed lepton doublet, with generation indices suppressed.
When $\sigma$ aquires a vacuum expectation value we can write
\begin{equation}
\sigma = \frac{1}{\sqrt{2}} (v_s + \rho + i \chi).
\end{equation}
Upon symmetry breaking, $\nu_R$ acquires a Majorana mass, $M = \sqrt{2}h_2 v_s$
and a Dirac mass with $\nu_L$, $m = h v / \sqrt{2}$ where $v/\sqrt{2} = 174$
GeV is the VEV of $H$. For $M \gg m$ the
low energy mass eigenstates are determined
by the seesaw relations \cite{seesaw}
$N = \nu_R +{\nu_R}^c + (m/M)({\nu_L}^c + \nu_L)$
and $\nu = \nu_L + {\nu_L}^c - (m/M)({\nu_R}^c + \nu_R)$
with $m_N \simeq M$ and $m_\nu \simeq m^2/M$. The
field $\rho$ has mass $m_\rho \sim v_s$ for self-couplings of
order 1, as is often assumed, and the massless majoron is $\chi$.

Because this model introduces only new singlet fields, it is only weakly
constrained.  If any of the light neutrinos are heavier than $\sim 100$
eV, they must decay fast enough so that neither they nor their
decay products overclose the Universe \cite{cmp2}. Furthermore, they must
decay fast enough to allow for sufficiently long late-time matter
domination, so that primordial density fluctuations can grow to form galaxies
\cite{pr}. The observed neutrino pulse from SN 1987A leads to a constraint on
the VEV, $v_s$ for neutrinos in the mass range from 100 eV to $\sim$ 30 MeV
\cite{cm}.  A massless majoron would also conflict with bounds on light
particle degrees of freedom \cite{sos} unless it decoupled early enough
\cite{bert}.  For the very small neutrino masses we will argue for below, the
majoron was always sufficiently decoupled.

\section{The Gravitationally Induced Bound}

The gravitationally-induced interactions for the singlet majoron model,
considered in \cite{ak}, were assumed to have the generic form
\begin{equation}
\lambda\frac{\sigma^5}{M_P}
\end{equation}
with the consequence that for $v_s > v$, a majoron mass
\begin{equation}
m_\chi \simeq \sqrt{20 \lambda} \left( \frac{v_s}{v} \right)^{3/2} {\rm keV}
\label{mchi}
\end{equation}
is generated.  Similarly, although we will not be interested in this
case, when $v_s < v$, a majoron mass of at least O(keV) will also be produced.
We anticipate that $\lambda \sim 1$ (there is no reason to assume otherwise),
but will nevertheless keep all $\lambda$-dependence explicit in our results.

An immediate consequence of majorons with a mass in the keV range is that they
would overclose the universe, if they were stable.  To see this, one must
estimate their relic abundance, $Y=n_\chi/n_\gamma$, given by the ratio of
particle species in equilibrium today to the number at the time they decoupled.
Above the phase transition where $\sigma$ gets its VEV, all the particles are
massless, so that majorons are in equilibrium at least until the critical
temperature $T_c$, which is of order $v_s$.  If this occurs above the weak
scale, but below the threshold for new SUSY or GUT particles, then the spectrum
of particles in equilibrium will just be all those of the Standard Model,
resulting in the dilution of the majoron number density by a factor $Y \simeq
(1/27)$ \cite{sos}. This dilution is sufficient to satisfy the constraints from
big bang nucleosynthesis \cite{wssok} since the the majoron at the time of
nucleosynthesis would only contribute the equivalent of $\sim$ 0.03 neutrinos
\cite{sos,bert}, well below the bound 0.3. However, the present density of
majorons becomes
\begin{equation}
\rho_\chi = \frac{1}{2} m_\chi Y n_\gamma
\end{equation}
which leads to the constraint
\begin{equation}
 m_\chi  <  340 {\rm\ eV} \left(\frac{1}{27Y}\right)
\left({\Omega h^2 \over 0.25}\right)
\end{equation}
and which for $\lambda\ga 0.01(v/v_s)^3$
is incompatible with the estimate in
(\ref{mchi}) \cite{ak}. If the majorons are kept in equilibrium to
lower temperatures, for example through the processes $N\leftrightarrow\chi\nu$
or $\nu\nu\leftrightarrow\chi\chi$, $Y$ will be larger and the conflict will
only be exacerbated.  Note that the bound cannot be evaded by assuming the
majoron decoupled at a higher temperature (resulting in a smaller value of
$Y$), since the majoron mass grows faster as a function of $v_s\sim T_c$ than
does $Y^{-1}$.  Thus majorons must decay.  But not only must they avoid
overclosing the universe, they must decay  early enough to allow for the
growth of density fluctuations \cite{pr}.

The recent measurements by COBE \cite{cobe} of the quadrupole moment in the
spectrum of primordial  density fluctuations, which is consistent with a flat
spectrum and an amplitude $\frac{\delta \rho}{\rho} \simeq 5 \times 10^{-6}$,
indicates the need for a substantial period of matter domination to allow for
the growth of these perturbations in order to form galaxies. Because of the
majoron mass scale, $m_\chi Y \ga 0.1$ keV, galaxy size perturbations will go
nonlinear only if the last epoch of matter domination is sufficiently long
\cite{pr,seealso}.  With such a small value of $\frac {\delta \rho} {\rho}$ it
is necessary to assume that majorons never dominate the overall density. This
implies that \cite{seealso}
\begin{equation}
\frac{m_\chi Y}{T_D} \la 9.0
\end{equation}
where $T_D$ is the majoron decay temperature. Relating $T_D$ to the
majoron lifetime gives the limit
\begin{equation}
\left(\frac{m_\chi Y}{1\rm\ keV}\right)
\left(\frac{\tau_D}{1\rm\ sec}\right)^{1/2} \la 1 \times 10^4.
\end{equation}
Then using
\begin{equation}
{\tau_D}^{-1} = \Gamma_D = \frac{m_\chi}{16\pi}~ \left(\frac{m_\nu}{v_s}
\right)^2
\end{equation}
one can obtain the bound \cite{ak} on $v_s$,
\begin{equation}
v_s \la 20 {\rm\ TeV} \left(\frac{m_\nu}{25\rm\ eV}\right)^{4/7}
\left(\frac{1}{27 Y}\right)^{4/7}
\lambda^{-1/7}
\label{vs}
\end{equation}
which is interesting in and of itself (note the weak dependence on
$\lambda$), as long as $m_\nu\la 25$ eV. In fact the latter condition {\it
must} be satisfied, because $m_\nu$ here refers to the heaviest neutrino which
can result from the decay $\chi\to\nu\nu$.  Such a neutrino, if heavier than
the cosmological bound of 25 eV, would itself have to decay into three lighter
neutrinos, $\nu\to 3\nu'$, with a majoron in the intermediate state.  However
the $\nu\nu'\chi$ coupling is highly suppressed \cite{sv}; at low energies
the Yukawa interactions of the singlet fields have the form
\begin{equation}
\frac{h_2}{\sqrt{2}} \chi \left({\bar N}XN -2 \frac {m}{M} {\bar
N}X\nu  + \frac {m^2}{M^2} {\bar \nu}X\nu + O\left(\frac {m^4}{M^4}\right)
{\bar \nu}X\nu'\right),
\label{coup}
\end{equation}
where $X= \rho+i\gamma_5\chi$. Because the rate is suppressed by the tiny
factor $(m/M)^{12}=(m_\nu/M)^6$, this decay mode is far too slow to vitiate the
25 eV bound on $m_\nu$.

\section{Sphaleron Induced Bounds}

We now turn to the baryon wash-out constraints. Any model with lepton
number violation faces the danger of baryon number wash-out as a result
of the combination of its own lepton number violating interactions
and the $(B+L)$-violating sphaleron interactions in the Standard Model.
This was first noticed by Fukugita and Yanagida \cite{fy2},
who showed that the presence of a lepton (or baryon) number violating
operator such as $\frac{h}{M}\nu\nu HH$ will erase any primordial baryon
asymmetry if these interactions are in equilibrium simultaneously with
sphaleron interactions. The condition that these interactions be out of
equilibrium translates into a bound on the neutrino mass
\cite{fy2,ht,nb}
\begin{equation}
m_\nu \la 50 {\rm\ keV} \left(\frac{100{\rm\ GeV}}{T_{BL}}\right)^{1/2}
\label{fuku}
\end{equation}
where $T_{BL}$ is the temperature at which a $B-L$ asymmetry was produced or
$\sim 10^{12}$ GeV (the maximum temperature for which sphaleron interactions
are in equilibrium), whichever is smaller.  This type of bound has been
extended to general baryon and lepton number violating operators in
\cite{cdeo12} where some general constraints on the majoron model were also
considered.

Another, complementary bound on neutrino masses can be obtained
considering the lepton number violating decays and inverse decays
$N \leftrightarrow HL$, that
inevitably accompany any nonrenormalizable operator such as indicated above.
The requirement that these processes are out of equilibrium has been
shown to imply
\cite{sonia}
\begin{equation}
m_\nu \la 10^{-3} {\rm\ eV}
\label{son}
\end{equation}
if the $B-L$ asymmetry is generated above $T\simeq M$. Remember that both
(\ref{fuku}) and (\ref{son}) can be evaded if the baryon asymmetry is
ultimately produced during or after the electroweak transition.

Without the bound (\ref{vs}) on the singlet majoron model,
it would be natural to assume that $v_s$ and $M$ are large. For
$T_{BL} < M \sim v_s$, the lepton-violating decays and inverse decays
would have gone out of
equilibrium before baryogenesis, evading the bound (\ref{son}).
Then one would be left with only the weaker constraint coming from
the dimension-5 operator (\ref{fuku}). Until a  nonvanishing neutrino mass is
measured, this puts no restriction on the scale of baryogenesis.  But with
the restriction (\ref{vs}) the story changes dramatically,
and the window of opportunity for producing the $B-L$ asymmetry
becomes very narrow.  With (\ref{vs}) the bound
from decays and inverse decays is the {\em only} one available as the
dimension-5 operators go out of equilibrium at $T \sim 10$ TeV, for
$m_\nu \simeq 25$ eV, and cannot be used at $T > v_s$.

The original derivation of the bound (\ref{son}) was incomplete in the sense
that the authors did not solve, nor even formulate the Boltzmann
equations for the asymmetries.
Because we are studying a theory with spontaneous symmetry breaking, where
the lepton-violating decays leading to (\ref{son}) only exist below
the critical temperature $T_c$ at which $v_s$ starts to develop,
and since we are
finding that $v_s$ must be near the weak scale,
it behooves us to reexamine in greater detail under what
conditions the validity of (\ref{son}) is generally assured.

Our basic assumption is that all the other processes proceed fast
compared to those that violate lepton number, so that at each instant
of time an equilibrium between various chemical
potentials is attained. The evolution of all
asymmetries is then controlled by the evolution of the individual lepton
asymmetries due to the decay processes $N \leftrightarrow HL$.

It is convenient to choose the mass eigenstates of the right-handed sector
as the two helicity states of the previously defined
Majorana field $N = \nu_R + {\nu_R}^c$, which
 does not mix with the light states until temperatures below the weak scale.
Above the critical temperature of the
phase transition  in the singlet sector these are precisely
the eigenstates of the conserved right-handed neutrino lepton
number.  Moreover it is convenient to work in the basis in which the
Majorana mass matrix is diagonal.  Since we are interested in
temperatures above the weak breaking scale, all the other fermions
remain massless.  As the weak symmetry eventually gets broken, the light
neutrinos gain a seesaw mass matrix whose diagonalized form is
\begin{equation}
m_\nu = U^T m^T \frac{1}{M} m U,
\label{seesaw}
\end {equation}
where $M$ is the heavy Majorana mass matrix, $m$ is the Dirac mass matrix
\begin {equation}
m \equiv hv/\surd 2,
\label{CKM}
\end{equation}
and $U$ is the neutrino mixing matrix that relates the flavor and mass
eigenstates.

The Boltzmann equation for the evolution of the $\nu_\ell$ number density,
due to the lepton number violating decays is given by
\vskip .1in
\begin{eqnarray}
(\partial_t - 3H)n_{\nu_\ell }  =  & {\displaystyle \int} & \! \prod_{i=1}^3
\frac{\textstyle \rd^3p_i}{\textstyle (2\pi)^32E_i}
(2\pi )^4\delta^4(p_1-p_2-p_3) \sum_j |{\cal M}|_j^2          \nonumber \\
  & \times & f_h(p_2,\mu_h) f_\nu(p_3,\mu_\ell ) e^{\beta E_1}  \nonumber \\
  & \times & \left\{ f_{j+}(p_1, \mu_j)(e^{-\beta \mu_j}
 - e^{\beta (-\mu_\ell -\mu_h)})  \right. \\
 & & \left. \mbox{}
 +  f_{j-}(p_1,-\mu_j)(e^{ \beta \mu_j} - e^{\beta (-\mu_\ell -\mu_h)})
\right\} \nonumber
\label{benl}
\end{eqnarray}
 where $\ell$
 is the
generation index.
The first term in the curly brackets in (\ref{benl}) comes from
$N_{j+}$ and the second from $N_{j-}$ decays, with the index $j$ running
over all heavy neutrino species and $H$ is the Hubble expansion rate of
the Universe. The maximal lepton number nonconservation below $T_c$ in
the decays $N_\pm \leftrightarrow H\nu_L,{H^*\nu_L}^c$  is reflected in the
fact that the helicity states $N_\pm$ decay  with equal total rate
to both allowed final states with opposite lepton number.
Indeed, the matrix element for the decay of $N_{j\lambda}$ is given by
\begin{equation}
|{\cal M}|_{j\lambda}^2 = |h_{j\ell}|^2 M_j
(p_j\cdot p_\nu \pm M_js_{j\lambda} \cdot p_\nu),
\label{matrix}
\end{equation}
where $s_{j\lambda}$ is the helicity four-vector of $N_j$ and $-(+)$ refers to
the $H{\nu_L}^c$ ($H^*\nu_L$) final state. The various decay channels
differ by the spin dependent part in (\ref{matrix}), but it is easy to
see that this contribution identically vanishes upon the phase space
integration.

Subtracting the Boltzmann equation for the number
density  of ${\nu_\ell }^c$'s
from that of $\nu_\ell$'s and linearizing in the chemical potentials,
one finds
\begin {equation}
(\partial_t - 3H)(n_{\nu_\ell }-n_{{\nu_\ell }^c})
 = - 4\:\frac{\mu_\ell + \mu_h}{T} \times  \sum_i h_{\ell i}^* \frac{
M_i(M_i^2+m_{\nu_\ell}^2-m_h^2)T}{64\pi^3} h_{i\ell} I_{i\ell}.
\label{asy1}
\end{equation}
The function
$I_{i\ell}$ can be evaluated exactly to yield
\begin{eqnarray}
I_{i\ell } = \int_1^\infty \rd u \frac{e^{ux}}{(1+e^{ux})^2}
\! &\ln \left( \frac {\textstyle \cosh (\alpha_\ell u + \gamma \sqrt{u^2-1})}
               {\textstyle \cosh (\alpha_h u - \gamma \sqrt{u^2-1})} \right.
\nonumber \\
& \times \left. \frac {\textstyle \sinh (\alpha_\ell u + \gamma \sqrt{u^2-1})}
             {\textstyle \sinh (\alpha_h u - \gamma \sqrt{u^2-1})} \right)
, \label{Fexa}
\end{eqnarray}
where $\alpha_\ell \equiv (M_i^2+m_{\nu_\ell }^2-m_h^2)/4M_iT$,
$\alpha_h \equiv (M_i^2+m_h^2-m_{\nu_\ell }^2)/4M_iT$
and $\gamma \equiv \lambda^{1/2}(M_i^2,m_{\nu_\ell }^2,m_h^2)
\linebreak
/4M_iT$, with $\lambda(x,y,z) \equiv (x-y-z)^2-4yz$. This is a
generalization of the inverse-decay rate used in \cite{fot}.   Note that the
$N_i$ chemical potentials do not appear in the equation (\ref{asy1}). This is
again due to maximal lepton number violation in the decays; should there be a
difference in the decay rates, the $N_i$ asymmetries would be nontrivially
coupled to $\nu_\ell $'s. In the present case the evolution of the $N_i$
asymmetry decouples and the $N_i$ asymmetries are driven to zero independently
of (and faster than) the asymmetries of the $\nu_\ell$'s. A subtlety in the
finite temperature calculation of the decay rate that deserves to be mentioned
is that the temperature dependent masses enter only through the kinematics and
the dispersion relation. In particular, the spinor functions must be taken
identical to the vacuum spinors, but with a modified dispersion relation
\cite{weldon}. In all our calculations we have used the approximate dispersion
relation $E = \sqrt{p^2+m^2(T)}$ where $m(T)$ is the appropriate temperature
dependent mass.

Equation (\ref{Fexa}) can be greatly simplified if one
substitutes the  Maxwell-Boltzmann approximation for the phase space
distributions. In general this is not legitimate, because then, for small
$M_i/T$, the  enhancement factor due to the singularity in the boson
distribution function at $E_h/T \simeq 0$ would be ignored In practice
however, this singularity is shielded by the temperature dependent mass of
the final state  particles and the MB approximation becomes effectively very
accurate.  We then find the approximate expression
\begin {equation}
I_{i\ell} \simeq \frac {\lambda^{1/2}( M_i^2,m_{\ell }^2,m_h^2 )}
{M_i^2} \cdot K_1\left(\frac {M_i}{T}\right),
\label{Fmb}
\end{equation}
where $K_1(x)$ is the modified Bessel function. This result coincides with
(\ref{Fexa}) to within few percent over the whole range of interest, and
becomes exact for large $M_i/T$.

The Higgs chemical potential $\mu_h$ can be solved in terms of the
leptonic chemical potentials from the equilibrium conditions imposed by
the fast weak processes. Following  \cite{ht}, and using
$\mu_\ell  = \mu_{\nu_\ell }$ due to fast weak interactions, we find
\begin{equation}
\mu_h = \frac{4}{21} \sum_{\ell=1}^3\mu_\ell.
\label{muh}
\end{equation}
Inserting this result into the equation (\ref{asy1}) and expressing the
chemical
potentials in terms of the asymmetries in the comoving volume, we obtain
\begin{equation}
L_{\nu_\ell } \equiv \frac{n_{\nu_\ell }-n_{{\nu_\ell }^c}}{s} \simeq
\frac {15}{4\pi^2g_*}\frac{\mu_\ell}{T},
\label{asymmetry}
\end{equation}
where s is the entropy density and $g_*$ is the number of relativistic degrees
of freedom.
We find the following coupled set of equations for the lepton asymmetries
\begin{equation}
\frac {\rd L_\ell}{\rd t} = - (L_\ell +
\frac{4}{21}\sum_{\ell '} L_{\ell '})\Gamma_\ell ,
\label{rate}
\end{equation}
with
\begin{equation}
\Gamma_\ell \equiv \frac{24}{T^3}\sum_i
{\vert h_{\ell i} \vert}^2T
\frac{(M_i^2+m_\ell^2-m_h^2)}{64\pi^3M_i}
\lambda^{1/2}( M_i^2,m_{\ell }^2,m_h^2 )
K_1\left(\frac{M_i}{T}\right).  \label{gamma}
\end{equation}
Eq.~(\ref{rate}) is valid as such for both neutrino and left handed
charged lepton asymmetries, which is why we replaced the subscript $\nu_\ell$
with $\ell$ in
(\ref{rate}) and (\ref{gamma}).  One should note that (\ref{rate}) allows for a
nontrivial equilibrium solution (some of the $L_\ell$'s nonzero) only if at
least one of the rates $\Gamma_\ell$ vanishes, or is small enough \cite{cdeo3}.
In the opposite case all $L_\ell$'s and hence all the other asymmetries as well
are driven to zero  with a rate controlled by (\ref{rate}). The baryon
asymmetry is related to the sum of the individual neutrino
 or left handed charged lepton asymmetries by
\begin{equation}
   B(t) = - \frac{4}{3} \sum_\ell L_\ell (t) .
\label{Basy}
\end{equation}
Given the initial asymmetries and the various couplings that
appear in the expressions for $\Gamma_\ell $, one could compute numerically
the baryon asymmetry as a function of time from equation (\ref{rate}).  For
the purpose of obtaining a bound on the neutrino mass, we can make the
following, most conservative, simplifying assumption about the parameters
involved in (\ref{rate}): ``the initial asymmetry corresponding to
the  {\it smallest} of rates $\Gamma_\ell $ is the {\it largest}.''
One should bear in mind that the fast electroweak processes enforce the
initial conditions  for the slow processes such that all initial asymmetries
are of the order as $L_\ell$'s.
The faster decaying asymmetries in eq.~(\ref{rate}) rapidly reach their fixed
point values
and it can be integrated exactly for the slowest one to yield:
\begin{equation}
L_\ell (t) = L_\ell (t_0)\exp \left(- \frac{33}{29}
\int_{t_0}^t \rd t
\: \Gamma_\ell (t)\right). \label{Lsolut}
\end{equation}
The bound on the neutrino mass arises from the requirement that the present
baryon asymmetry exceeds the known lower limit of $B_{min} \simeq 4\times
10^{-11}$, which in terms of the lepton asymmetries becomes
\begin {equation}
\left\vert \sum_\ell^3 L_\ell (t_{EW})\right\vert > \frac {4}{3} B_{min},
\label{b1}
\end{equation}
where the time $t_{EW}$ refers to the time of electroweak phase
transition. With our conservative assumption we then obtain the bound
\begin{equation}
{\left( \int_{t_0}^{t_{EW}} \! \rd t \: \Gamma_\ell \right)}_{min} \la
\frac{29}{33} \ln\left(\frac{L_\ell^0}{4\cdot 10^{-11}}\right) \simeq 17,
\label{b2}
\end{equation}
using in the last step a reasonable maximum value of
${L_\ell^0}_{max} \sim \epsilon n_\gamma /s \la 10^{-2}$ for the initial
asymmetry, where $\epsilon$
is some generic $CP$-asymmetry parameter \cite{fot}.  Up to now our
discussion has been quite general and eq.~(\ref{b2}) is valid for either
explicit or spontaneous symmetry breaking. Assuming that the temperature
dependent mass of the $N_i$ is given by
\begin{equation} M_i^2(T) =
M_{0i}^2(1-(T/T_c)^2), \label{Nmass}
\end{equation}
we can
write the integral appearing in (\ref{b2}) in the spontaneous symmetry breaking
 case as
\begin {equation}
\int_{t_c}^{t_{EW}} \! \rd t \: \Gamma_\ell \simeq 284 \sum_i {m^*}_{\ell i}
\frac{1}{M_i}{m}_{i\ell} \times I
\label{integral}
\end{equation}
where
\begin{equation}
I \equiv \int_0^{z_{EW}} \! \rd z
\frac {z^4}{\sqrt{z^2+(\frac{M_{0i}}{T_c})^2}} f(z) K_1(z),\label{ii}
\end{equation}
with $z \equiv M_i(T)/T$, $f(z) =
(1+r_\ell^2-r_h^2)\lambda^{1/2}(1,r_\ell^2,r_h^2)\theta(1-r_\ell-r_h)$ and
$r_\alpha \equiv m_\alpha (T)/M_i(T)$. The
explicit symmetry breaking case can be obtained by  taking the
limit $t_c \rightarrow 0$, $T_c \rightarrow \infty$.
In the singlet majoron model the various temperature dependent masses that we
use in our calculations are given by
\begin{eqnarray}
M_i^2(T) &\equiv & 2 h_{2i}^2 v_s^2(T) \qquad ; \qquad v_s^2(T) =
v_s^2(1-T^2/T_c^2) \nonumber \\
m_\ell^2(T) &=& \frac{1}{32}(3g^2+{g'}^2)T^2 \\
m_h^2(T) &=& \frac {1}{16}(3g^2 +{g'}^2 + 4h_t^2 + 8\lambda_H +
\frac{4}{3}\lambda_{sH}) T^2 - \lambda_H v^2 -
\frac{1}{2}\lambda_{sH}(v_s^2-v_s^2(T)),
\nonumber
\label{masses}
\end{eqnarray}
where $g$ and $g'$ are the usual weak and $U(1)_Y$
couplings, $h_t$ is the top quark Yukawa coupling,
$\lambda_H$, $\lambda_s$, $\lambda_{sH}$, $v$ and $v_s$ are the quartic scalar
couplings and VEV's defined in the equation (\ref{effpot}) and $T_c$ is the
$(B-L)$-symmetry breaking temperature given by
\begin{equation}
T_c =
v_s\left({6\lambda_s +3\lambda_{sH}(v/v_s)^2\over
2\lambda_s+\lsh+\vec{h}_2^2}\right)^{1/2},
\label{tc}
\end{equation}
where $\vec{h}^2 \equiv \sum_j h_{2j}^2$.  The expression for $m_h(T)$
in (\ref{masses}) assumes that for temperatures $T_c > T > T_{EW}$,
the field is located in the
temperature dependent asymmetric miminum in the singlet direction with the
electroweak symmetry still unbroken, $\langle |\sigma | \rangle = v_s(T)/2$ and
$\langle |H| \rangle = 0$.
We have computed the integral $I$ for a wide range of parameters and find
that it is essentially constant $I \simeq 4$, for $M_i > m_h$.
  The constant behaviour arises because the integrand $\sim z^3K_1(z)$ peaks
for rather large values of $z \simeq 3$ so that the effects of final state
masses and the $M_{0i}/T_c$ dependence are strongly suppressed.  Thus the final
bound arising from (\ref{b2}) is essentially the same for either spontaneous or
explicit breaking. With a constant $I$ the LHS of the equation (\ref{b2})
becomes proportional to the quantity $(m^\dagger M^{-1}m)_{\ell \ell}$.

Our main result now comes from combining  (\ref{b1}) and
(\ref{b2}), using the constant value of $I$ just determined:
\begin{equation}
\min_\ell \: (m^\dagger M^{-1}m)_{\ell\,\ell}
 < 8 \times 10^{-4} \ln \left(\frac{L_0}{4\cdot10^{-11}}\right)
{\rm\ eV\ } \simeq
10^{-2}{\rm\ eV\ }, \label{final}
\end{equation}
where again, the maximum value for $L_0$ was used in the last step.
Notice that the careful treatment leading to eq.~(\ref{final}) has yielded a
somewhat less restrictive bound on the scale of the seesaw neutrino masses,
$m^2/M$, than previous investigations.  But another interesting feature of
(\ref{final}) is that, unlike previous derivations, it does not
give any direct bound on the masses of the neutrinos, because $m^\dagger
M^{-1}m$ is different from the mass matrix $m^T M^{-1}m$.  For example,
consider just two generations and assume that $M^{-1/2}m$ is proportional to
the Pauli matrices $\sigma_1+\sigma_2$. Then $m^\dagger M^{-1}m$ is diagonal
and nonzero, but the mass matrix vanishes identically, and thus the
bound (\ref{final}) is not directly related to the masses.
 Such a situation could arise
either due to fine-tuning of the mass matrices or (approximate) lepton flavor
symmetries. Otherwise one generically expects $m^\dagger M^{-1}m$ to be
related  to the masses and mixings via
\begin{equation}
(m^\dagger M^{-1}m)_{\ell\,\ell} \sim \sum_{i=1}^3 m_{\nu_i}|U_{i\ell}|^2,
\label{mdMm}
\end{equation}
as can be seen by writing $M^{-1/2}m = V^\dagger D U$, where $D$ is diagonal,
and assuming that $V$ is either close to the identity, or has off-diagonal
elements of the same order as those of $U$.  Then $m^\dagger M^{-1}m =
U^\dagger |D|^2 U$, whereas the mass matrix is $U^T D V^T V D U$.  If $V^TV = 1
+ \epsilon$, then (\ref{mdMm}) is correct to lowest order in $\epsilon$.  If
$\epsilon$ is large but the off-diagonal elements of $U$ are also large, then
(\ref{mdMm}) will again be approximately true, since it is dominated by
the largest eigenvalue of $D$, that is, the heaviest neutrino mass, regardless
of the generation index $\ell$.

If either of the reasonable assumptions leading to eq.~(\ref{mdMm}) holds, we
can combine the baryon-preservation bound (\ref{final}) with the majoron decay
bound (\ref{vs}) to obtain what is potentially a much more stringent limit on
the majoron scale than was given previously in eq.~(\ref{vs}).
Eq.~(\ref{final}) tells us that the mass of the heaviest neutrino $\nu_h$ into
which the majoron can decay should satisfy $m_{\nu_h} < 10^{-2}U_{h\ell}^{-2}$
eV, where $\ell$ is the neutrino flavor whose conservation is most weakly
violated.   Then (\ref{vs}) becomes
\begin{equation}
v_s \la {\rm max}\left(200 {\rm\ GeV} (27 Y)^{-4/7}U_{h\ell}^{-8/7}
\lambda^{-1/7},\ v \right).
\label{us}
\end{equation}
Note that we cannot obtain a bound for $v_s < v$, since sphaleron interactions
will have decoupled by then.

A further application of our bound is a constraint on the Yukawa couplings of
the right-handed neutrinos to the Higgs doublet, about which no information has
hitherto been available.  Rewriting the mass matrices in terms of Yukawa
coupling matrices and using the constraint (\ref{us}) on $v_s$ we obtain
\begin{equation}
\min_\ell \: \sum_i \frac{|h_{i\ell}|^2}{{h_2}_i} < 6\cdot 10^{-14}{\rm
max}\left(0.8(27 Y)^{-4/7}U_{h\ell}^{-8/7} \lambda^{-1/7},\ 1\right).
\label{ours2}
\end{equation}
Even if the gravitationally induced effects (parametrized by $\lambda$) or the
mixing $U_{h\ell}$ are very small, this is quite a strong result, for it says
that either all the Yukawa couplings $h_{i\ell }$ are even smaller than that of
the electron or there exist large hierarchies within $h_{i\ell}$ with some of
the elements being extremely small; $h_{i\ell} \la 10^{-7}$.

The dramatic results (\ref{us},\ref{ours2}) apply in the case of generic mass
matrices for the neutrinos, but in model building it is often interesting to
consider the nongeneric case where some combination of flavors $\Lt$ is
approximately conserved, for example $\Lt = L_e - L_\mu + L_\tau$. If $\Lt$ is
only weakly violated, then the decay vertices we have considered have only very
slow $\Lt$-violating channels, since the heavy decaying neutrinos $N$ have
pseudo-Dirac masses of order $M$ (which almost conserve $\Lt$) rather than
Majorana masses that would violate $\Lt$ maximally.  The decays are therefore
predominantly of the form $N\to H\nu$ and $\overline N\to H^*\bar\nu$. The
fastest $\Lt$ violating process are $N$-$\overline N$ oscillations coming from
the small Majorana mass terms $\mu$ \cite{cline}, followed by the
$\Lt$-conserving decays. These are faster than any oscillations ocurring in the
light neutrino sector because the mass splittings, hence the oscillation rate,
are much larger for the heavy neutrinos.  To be exact, we would have to write
the Boltzmann equation also for the asymmetry in $N$ and combine it with that
for $\nu$ in order to track the asymmetry in $\Lt$.  But roughly, we can
account for this effect by multiplying the previously computed decay
probability by the probability of an oscillation, given by
\begin{equation}
P = \sin^2(2\theta)\sin^2\left({\mu M\over 2E\Gamma}\right)
\label{prob}
\end{equation}
where for pseudo-Dirac neutrinos the mixing is maximal ($\sin^2(2\theta)=1$),
and $\Gamma$ is the rate of interactions of $N$ which will damp the
oscillations if they are much faster than the oscillation rate $\delta M^2/2E
=\mu M/E$.  $\Gamma$ is given by the sum of rates $\sum_\ell\Gamma_\ell$ which
we computed previously, eq.~(\ref{gamma}). For very fast oscillations the
second $\sin^2()$ factor averages to $1/2$, but if $\mu$ is small then our
bound (\ref{final}) starts to disappear.  The condition for (\ref{final}) to be
valid in the case of an approximate symmetry is therefore that $\mu M\ga
T\Gamma$ at temperatures of order $M$, or in terms of the dimensionless
paramter $\mu/M$ that quantifies the small departures from $\Lt$ symmmetry,
\begin{equation}
{\mu\over M}\ga {1\over\pi^3}\sum_{i\ell}|h_{i\ell}|^2\sim {\sum_i
m_{\nu_i}\over \pi^3(v^2/M)}.
\label{breaking}
\end{equation}
Therefore even in the extreme case of $M = 100$ GeV and $m_{\nu_3}=10$ MeV,
a very small symmetry breaking $\mu/M\sim 10^{-6}$ would be sufficient for our
results to still apply.

\section{Caveats}

Finally, we discuss some special situations, where our bounds (\ref {us})
and (\ref{ours2}) could be avoided. One such situation is the case in
which the baryon asymmetry is actually generated below  $T_c$.  Since the
wash-out due to the decays is not effective for $T\ll M_i(T)$,
the asymmetry generated at $T_{BL}$ is safe if $T_{BL} \ll T_c$. However, due
to
the bound (\ref{vs}) on $v_s$ with $T_c \sim v_s$ this should be seen as an
extremely tight constraint on the scale $T_{BL}$!

Another possibility is that the baryon asymmetry is generated at some high
scale, and the electroweak symmetry breaks before the $B-L$ symmetry even
though $v_s > v$. If this was the case, then there never was a period where the
sphaleron  interactions and the lepton number violation were in equilibrium
together and the bound (\ref{mdMm}) would not be valid. To see how generic this
is, we must compare the critical temperature for the electroweak breaking (at
the field origin) with the $(B-L)$-symmetry breaking scale $T_c$ (\ref{tc}).
We find
\begin{equation} T_{EW} =
v \left({12\lambda  +6\lambda_{sH}(v_s/v)^2\over
6\lambda+\lsh+3r}\right)^{1/2},
\label{tew}
\end{equation}
where $r = (2M_W^2+M_Z^2+2m_t^2)/v^2$.  Note that if the $B-L$ symmetry is
indeed broken first, the temperature (\ref{tew}) is {\em not} the true
electroweak  breaking temperature, since the field shifts from the origin after
$(B-L)$-breaking.  The parameters of the quartic couplings in the scalar
potential must always satisfy the relation $\lambda_{sH} < 4\lambda \lambda_s$
in order for {\em both} symmetries be broken; otherwise there would be no light
neutrino masses at low scales. Moreover, the requirement that the one-loop
corrections do not render our vacuum  unstable introduces a bound on the Yukawa
couplings $h_{2i}$.  This bound can be naively estimated by computing the
coefficient of the $|\sigma|^4 \ln |\sigma|$ term and requiring that this
coefficient be nonnegative: ($\frac{5}{4}\lambda_s^2 + \frac{1}{16}
\lambda_{sH} - \sum_j h_{2j}^4) > 0$.  One can then show that the
aforementioned constraints with $v<v_s$ and $T_{EW}>T_c$ would be satisfied
only in a small region of parameter space, where
 $\lambda_s \ll\lambda_{sH} \ll \lambda \sim {\cal O}(1)$.

There are two more subtle cases where the bound (\ref{final}) could be
weakened, related to the computation of the integral $I$ in (\ref{ii}).
First, it was assumed that the parameter $M_{0i}/T_c$ is not excessively large.
Using the vacuum stability bound and the bound on the quartic couplings,
one can show that $M_{0i}/T_c \ga$ some number $\alpha$ would require
$\sqrt{3} \alpha (1+(\frac{\lambda}{\lambda_s})^{1/2})^{-1/2} \la h_2 \la 2
\lambda_s$.  This inequality has solutions for $\alpha \ga 1$, i.e., only for
nonperturbative values of couplings
$\lambda, \lambda_s, h_2 \ga {\cal O}(1)$.
A more severe problem arises if $M_{0i}/T_c$ is excessively small.
As noted earlier, the integral in (\ref{ii}) gets its largest
contribution from the region $z \equiv M_{0i}(T)/T \simeq 3$.
For small $M_{0i}/T_c$ this peaking occurs at temperature
$T_p \simeq M_{0i}/3$. It is always possible to make $M_{0i}$ much less than
$T_c$ merely by letting $h_2$ become small. When $M_{0i} < m_h$ our bound
scales roughly as $h_2^{-1}$.
However, when $v_s$ is small, a very small
value of $h_2$ is rather unattractive anyway, since the mass $M_{0i}$
becomes very small, rendering the seesaw mechanism less natural.

\section{Conclusions}

While it is clear that small neutrino masses are an interesting and potentially
necessary extension to the standard model, the origin of these masses
 remains unclear.  Many of the possibilities involve lepton number
violating Majorana masses, as is the case in the singlet majoron model
discussed here. The combination of lepton number violation in conjunction
with the $B+L$ violation  in the standard model leads to the grave possibility
of washing out {\it any} primordial baryon asymmetry.  Thus, the survival of
the
baryon asymmetry enables one to constrain models such as the singlet majoron
model (unless, of course, the baryon asymmetry is produced very late). We were
thus able to strengthen the bounds of ref.~\cite{ak}.  We have shown under
general circumstances that at least one neutrino mass is constrained to have a
value $m_\nu \la 10^{-2}$ eV and the neutrino Yukawa coupling is constrained to
$h \la \rm{few} \times 10^{-7}$. The singlet VEV is correspondingly constrained
by $v_s \la v$.  While the majoron model is still viable with a low value of
$v_s$, models with a large VEV must be carefully tuned so as to avoid the
constraints discussed here as well as in \cite{ak}.

\vskip .4in

\noindent{ {\bf Acknowledgements} } \\
\noindent  We would like to thank B. Campbell, S. Davidson, M. Davis,
S. Nussinov and S.~Paban for useful discussions.  This work was supported in
part by  DOE grant DE-AC02-83ER-40105. The work of KAO was in addition
supported by a Presidential Young Investigator Award.
}}

\end{document}